\documentstyle[12pt]{article}
\input epsf
\begin{document}
\begin{titlepage}
\title{On the Phase-Space Volume of Primordial Cosmological Perturbations}
\author{J. Lesgourgues$^1$, David Polarski$^{1,2}$ and Alexei A. Starobinsky$^a$\\
\hfill \\
$^1$~{\it Lab. de Math\'ematique et Physique Th\'eorique, EP93 CNRS}\\
{\it Universit\'e de Tours, Parc de Grandmont, F-37200 Tours (France)}\\
\hfill\\
$^2$~{\it D\'epartement d'Astrophysique Relativiste et de Cosmologie},\\
{\it Observatoire de Paris-Meudon, 92195 Meudon cedex (France)}\\
\hfill \\
$^a$~{\it Landau Institute for Theoretical Physics}, \\
{\it Kosygina St. 2, Moscow 117334 (Russia)}}

\date{\today}
\maketitle

\begin{abstract}
We show how to determine the typical phase space volume $\Gamma$ for 
primordial gravitational waves produced during an inflationary stage, 
which is invariant under squeezing.  
An expression for $\Gamma$ is found in the long wavelength regime. 
The quasi-classical entropy of a pure vacuum initial state defined as the 
logarithm of $\Gamma$ modulo a constant remains zero in spite of the 
generation of fluctuations (creation of real gravitons).
\end{abstract}

\end{titlepage}

\section{Introduction}

As well known, the inflationary scenario offers an elegant solution to some 
of the outstanding problems of the big bang cosmology like the horizon and 
the flatness problems. In this scenario primordial scalar (adiabatic) 
perturbations~\cite{hsg} and primordial gravitational waves~\cite{st79} are
produced as a 
consequence of the anavoidable quantum fluctuations of some field(s).
The scalar (adiabatic) fluctuations eventually lead to the formation of stars,
galaxies and the large scale structure in the Universe, while gravitational 
waves (GW) forms a stochastic waves background that could be in principle 
presently observable.
An intriguing property of the produced fluctuations is their quantum origin. 
It turns out that the inflationary scenario provides us, in addition to the 
above mentioned achievements, with a physical system, namely the primordial 
fluctuations, undergoing a quantum to classical transition solely as a result 
of its dynamics~\cite{david96}. 
We will show that the effective volume in phase-space $\Gamma$ occupied by the 
primordial GW generated during inflation is constant, it is not the naive 
product of variances of the canonically conjugate 
variables $y$ and $p$. The latter definition yields a large phase-space
volume for 
large squeezing and we show that it corresponds to a complete loss of the 
classical correlation between the quantities $y$ and $p$ for large squeezing. 

The determination of the typical phase space volume of cosmological 
perturbations
is closely connected with the question of their entropy. In particular, 
we use the semiclassical definition for the entropy $S$ yielding 
$S=\ln \Gamma + \beta,~(\beta={\rm constant})$.
Note that generation 
of fluctuations in the inflationary scenario may be considered as a 
particular case of particle creation in a Friedmann-Robertson-Walker (FRW) 
isotropic cosmological
model. Pioneering papers on the entropy of particles created in a FRW model
~\cite{hu} define the entropy through the average number of created particles
neglecting phase correlation between them at late times (this just results
in the transformation of a pure quantum state into a mixed one).   
After that, there were numerous attempts~\cite{pro,gasperini} to apply
this approach to cosmological perturbations and to find their entropy 
either by neglecting the phase correlations or using another ad hoc coarse 
graining procedure. Though coarse graining and neglection of phases is 
possible in the case of creation of massive particles which become
thermalized at very early stage of the evolution of the Universe,
it is not justified in the case of cosmological perturbations which are
massless (as GW) or almost massless (as the inflaton field) and very weakly
interacting. We have stressed already that the actual phase space volume of 
perturbations is much less than the one that would arise as a result of such 
a coarse graining~\cite{david96} and we want now to consider this question in 
more detail.

\section{Evolution of primordial quantum gravitational waves}

We will consider for simplicity primordial GW produced during an 
inflationary stage~\cite{david96}. They originate from vacuum fluctuations 
of the quantized tensorial metric perturbations. Each polarization state 
$\lambda$, where $\lambda=1,2$, and the
polarization tensor is normalized to $e_{ij}e^{ij}=1$, has an 
amplitude $h_{\lambda}$ given by
\begin{equation}
h_{\lambda}=\sqrt{32\pi G}\phi_{\lambda}
\end{equation}
where $\phi_{\lambda}$ is a real massless scalar field. We will omit in the 
following the polarization subscript $\lambda$ and introduce the rescaled 
field $y\equiv a\phi$, $a$ being the scale factor. 
Let us consider the dynamics of a real massless scalar field on a flat FRW 
universe. Actually, the field can be massive and all the arguments presented 
here will go through. As a result of the coupling with the gravitational 
field, which plays here the role of an external classical field, the state 
$|0,~\eta \rangle$ which is the vacuum state of the field at some 
given initial time $\eta_0$ will no longer be 
the vacuum at later time. Indeed field quanta are produced in pairs with 
opposite momenta, we get a two-mode squeezed state. Though the 
state at late 
time is no longer annihilated by the annihilation operators $a({\bf k})$, it 
is still annihilated by the following time-dependent operator
\begin{equation}
\Bigl \lbrace y({\bf k},\eta_0)+i\gamma_k^{-1}(\eta)p({\bf k},\eta_0)
\Bigr \rbrace|0,\eta\rangle_S=0~.\label{eq2}
\end{equation}
Therefore the state at late time is still Gaussian, its wave functional in 
the infinite volume case (continuous $\bf k$'s) being 
%an (infinite) product of functions 
\begin{eqnarray}
\Psi[y({\bf k},\eta_0),y(-{\bf k},\eta_0)] 
&=& \sqrt{{\cal N}_k}\exp \left(-\frac{1}{2}\int d^3{\bf k}~\gamma_k~~
y({\bf k},\eta_0)~y(-{\bf k},\eta_0) \right) 
%&=& \sqrt{N_k}\exp \left(-\frac{|y({\bf k},\eta_0)|^2} {2|f_k|^2}
%\lbrace 1-i2F(k)\rbrace \right)\label{Psi}
\end{eqnarray}
%for each pair ${\bf k},-{\bf k}$
where ${\cal N}_k$ is a normalization coefficient. 
%The time dependence of $\Psi$ is through $f_k,~F(k),~{\rm and}~N_k$.
We note the persisting symmetry under reflections in $\bf k$-space, it is 
present in the initial state and the evolution doesn't spoil it, we get a 
two-mode squeezed state.

The dynamics of the 
system is particularly transparent in the Heisenberg representation. We have,
using the field modes $f_k(\eta)$ with $\Re f_k\equiv
f_{k1}$ and $\Im f_k\equiv f_{k2}$, $f_k(\eta_0)=1/\sqrt{2k}$,
(we adopt a similar notation for all quantities)
\begin{eqnarray}
y(\bf k,~\eta)
&\equiv& f_k(\eta)~a({\bf k},\eta_0)+f_k^*(\eta)~
a^{\dag}(-{\bf k},\eta_0)\nonumber\\
&=&f_{k1}(\eta)~~e_y({\bf k})-f_{k2}(\eta)~~e_p({\bf k})
\label{yk}
\end{eqnarray}
and the momentum modes $g_k(\eta)$,~ $g_k(\eta_0)=\sqrt{{k\over 2}}$,
\begin{eqnarray}
p(\bf k,~\eta)
&\equiv& -i\bigl\lbrack g_k(\eta)~a({\bf k},\eta_0)
-g_k^*(\eta)~a^{\dag}(-{\bf k},\eta_0)\bigr\rbrack\nonumber\\
&=&g_{k1}(\eta)~~e_p({\bf k})+ g_{k2}(\eta)~~e_y({\bf k}).\label{pk}
\end{eqnarray}
The time independent operators $e_p({\bf k})$, resp.$e_y({\bf k})$, satisfy 
\begin{equation}
\langle e_y({\bf k})~e_y^{\dagger}({\bf k}')\rangle=
\langle e_p({\bf k})~e_p^{\dagger}({\bf k}')\rangle=
\delta^{(3)}({\bf k}-{\bf k}')
~~~~~~e_{y,p}^{\dagger}({\bf k})=e_{y,p}(-{\bf k})~. 
\end{equation}
They obey the commutation relations 
\begin{equation}
[e_i({\bf k})~,~e_j^{\dagger}({\bf k}')]=2i~\delta_{ij}~\delta^{(3)}
({\bf k}-{\bf k}')~,~~~~~~i,j=y,p~.\label{com}
\end{equation} 
The operators $e_y$ and $e_p$ correspond to standing, not running 
quantum waves.

The canonical commutation relations (\ref{com}) express the fact that the 
quantities $e_y,~e_p$ are operators and that the system is a 
quantum mechanical one.
Furthermore, 
\begin{equation}
\gamma_k=\frac{1}{2|f_k|^2}(1-i2F(k))~~~{\rm with}~~~ 
F(k)=f_{k1}~g_{k2}-f_{k2}~g_{k1}~,
\end{equation}
hence at $\eta=\eta_0$, we start with a minimum uncertainty wave function.
The field modes obey the Klein-Gordon equation
\begin{equation}
f''_k -\frac{a''}{a}~f_k + k^2~f_k = 0~,\label{KG}
\end{equation}
and are constrained to satisfy the condition
\begin{equation}
2 ( f_{k1}g_{k1} + f_{k2}g_{k2} ) = 1~,
\end{equation}
which can be viewed as either the Wronskian condition for (\ref{KG}) or the 
commutation relations imposed by canonical quantization.
We note that special interest is attached to the quantity 
$B(k)\equiv f^*_k g_k$, with 
\begin{eqnarray}
\Re B(k) &=& \frac{1}{2}~,\\
\Im B(k) &=& F(k)~.
\end{eqnarray}
The evolution of the system is conveniently parametrized by the squeezing 
parameter $r_k$, the squeezing angle $\varphi_k$ and the phase $\theta_k$
~\cite{schum}.
A crucial property is that for large squeezing, $|r_k|\gg 1$, the phase and 
squeezing angles do not evolve independently of each other, we can impose 
instead with a proper choice of the initial conditions~\cite{david96}
\begin{equation}
\theta_k+\varphi_k \simeq 0~.\label{phase}
\end{equation}

Wavelengths much larger than the Hubble rudius at the end of the 
inflationary stage are in a WKB quasi-classical state corresponding to 
$|F(k)|=\frac{1}{2}\sin 2\phi_k \sinh 2r_k\gg 1,~|r_k|\gg 1$, $f_k$ almost 
real and $g_k$ almost imaginary, and use has been made of 
(\ref{phase}) for the last property.  
We then get for the primordial GW field modes
\begin{eqnarray}
\sqrt{2k} f_k &\simeq& \cos \varphi_k~{\rm e}^{r_k}~,~~~~~~~
\sqrt{2k}~|f_k| \gg 1~,\label{f}\\
\sqrt{\frac{2}{k}} g_k &\simeq& \sin \varphi_k~{\rm e}^{r_k}\approx 0~.
\end{eqnarray}
The primordial GW dynamics can then be 
described by two real functions, namely $r_k$ and $\varphi_k$ or equivalently 
$f_{k1}$ and $g_{k2}$, instead of three.
In the limit $|r_k|\to \infty$, the dynamics of the system corresponds to a 
{\it classical stochastic} process: on one hand,
$y({\bf k}, \eta)=f_k(\eta)~e_y({\bf k})\equiv y_{cl}({\bf k})$ 
where $f_k$ resp. $g_k$, is real resp. imaginary, while $e({\bf k})$ is 
a time-independent Gaussian stochastic function of ${\bf k}$ with zero 
average and unit dispersion;
on the other hand, the Fourier transform $p({\bf k},\eta)$ of the momenta 
take their classical values $p_{cl}({\bf k},\eta)$
\begin{equation}
p_{cl}({\bf k},\eta) = -i~g_k~e_y({\bf k})= -i~\frac{g_k}{f_k}~y({\bf k},\eta)
\label{pcl}
\end{equation}
for each realization of the stochastic field $y({\bf k},\eta)$ and 
$y_{cl}$, resp. $p_{cl}$, is the {\it classical} amplitude, resp. 
momentum, for the corresponding initial conditions when the decaying mode 
is negligible. A generic outcome of the inflationary stage is the very high 
squeezing of the primordial GW on cosmological scales till after the last 
Hubble radius crossing.

\section{The phase-space volume $\Gamma$}

What is the typical volume $\Gamma$ occupied by the GW in phase-space?
This question is closely connected to the problem of the entropy as might be 
expected. 
It is instructive to look at it in the Heisenberg 
representation using eqs.~(\ref{yk},\ref{pk}). According to these equations, 
in the limit that our system is equivalent 
to a classical stochastic process, the volume occupied in phase-space  
should has measure zero! Indeed, while the amplitude is typically found 
inside a domain $\delta y$ which is of order of $\Delta y$, the variance of 
$y$, for 
each value of $y$ there corresponds only one momentum $p_{cl}(y)$. 
This can be expressed mathematically in the following way. If we define 
the conditional probability density ${\cal P}(p,~\eta~|~y_1,~\eta_1)$ 
for a certain value $p$ of the momentum $p$ at time $\eta$ provided the 
amplitude has a value $y_1$ at time $\eta_1$, we find for $|r_k|\to \infty$
\begin{equation}
{\cal P}_{cl}(p,~\eta~|~y,~\eta)= \delta~(p-p_{cl}(y))~,\label{Pcl} 
\end{equation} 
with $p_{cl}(y)\equiv \frac{g_{k2}}{f_{k1}}~y$. This is graphically 
illustrated in Fig.1 where the thin solid circle represents the initial 
vacuum state, the strongly elongated ellipse is the squeezed vacuum state, and
the straight line - the major axis of the ellipse - just gives the 
approximation of the quantum GW by a classical GW with 
a stochastic amplitude, a fixed phase and a probability distribution in 
phase-space given by (\ref{Pcl}).
It is crucial to properly 
take into account the stochasticity 
of the operators $y$ and $p$ as expressed by the correlation (\ref{pcl}) 
between these operators. Note that this analysis does not depend on a 
specific squeezed state, only the dynamics of the field modes, namely the 
vanishing of the decaying mode given by the second term in eq. (\ref{as})
below, is taken into account. Therefore the 
conclusions reached apply to more general squeezed states than just the 
squeezed vacuum and it could apply as well to other squeezed 
systems possessing a growing and a decaying mode~\cite{julien96}.
\par
Equation (\ref{Pcl}) is in sharp contrast to the phase space volume (we 
consider here and below our system to be enclosed in a finite volume)
\begin{eqnarray}
\Delta^2 y~\Delta^2 p 
&=& \langle y~y^{\dagger}\rangle~\langle p~p^{\dagger}\rangle = F^2(k)+\frac{1}{4}\nonumber\\
&=& 4~\Delta y_1~\Delta p_1~\Delta y_2~\Delta p_2.
\end{eqnarray}
The latter tends to $|F(k)|$ for
$|F(k)|\gg 1$, the fact that it becomes very large just reflects 
the semiclassical limit of the system. It implies in particular that in the 
Schr\"odinger representation the probability density 
$\rho(y_0,~\eta)=|\Psi|^2$ is spreading (in amplitude space) and that 
it moves along classical trajectories. 
Taking the quantity $\Delta^2y~\Delta^2p$ for the phase-space volume occupied 
by the system corresponds to the case 
where, for given $y$ inside $\Delta y$ we neglect information, which is 
available as a result of large squeezing,
about the different momenta inside $\Delta p$. We will come back to this 
important point at the end of this article.
In other words, with this definition one is not taking into account the true 
origin of the stochasticity of $y({\bf k},\eta)$ and $p({\bf k},\eta)$ 
expressed by (\ref{yk},\ref{pk}).
From the above discussion it turns out that, for $|r_k|\gg 1$, an ansatz 
of the kind
\begin{equation}
\Gamma\sim \Delta^2 y~\Delta^2 P~~~~~{\rm where}~~~~~
P\equiv p-\frac{g_{k2}}{f_{k1}}~y~, 
\end{equation}
would be more adequate as it possesses the correct behaviour in the limit 
$|r_k|\to \infty$. We shall now show that 
this choice comes naturally out when use is made of the Wigner function. 
\par
Indeed, in order to find $\Gamma$ one needs to consider a probability density 
in phase-space and not just in amplitude space.  
Of course, a quantum system does not (and cannot) possess a true probability 
density in phase-space, one can at best construct some candidates which have 
basic desired properties. The Wigner function~\cite{wigner} is such a 
well-known example. 
It is not positive definite in general but it has this property for any 
initial coherent state and in particular for an initial vacuum state as 
these are Gaussian states. Therefore, we will take the Wigner function as 
the probability density of our system in phase-space. 
The following Wigner function is obtained for any initial coherent state 
(the subscript $0$ applies to quantities at time $\eta_0$ and we insert here 
$\hbar$ back)
\begin{eqnarray}
W & = & N_k^2~e^{-\frac{|y_0-\langle y \rangle|^2}{|f_k|^2}}~
\frac{|f_k|^2}{\pi\hbar^2}~e^{-\frac{4|f_k|^2}{\hbar^2}
|\frac{F(k)}{|f_k|^2}(y_0-\langle y \rangle)-(p_0-\langle p \rangle)|^2 }\\
& \stackrel{|r_k| \rightarrow \infty}{\longrightarrow} & 
|\Psi|^2~\delta~\left( p_{01}-\frac{F(k)}{|f_k|^2}~y_{01}\right)
~~\delta~\left( p_{02}-\frac{F(k)}{|f_k|^2}~y_{02} \right)~.\label{wig}
\end{eqnarray}
Note that this Wigner function is calculated with the exact wavefunction
~\cite{anderson}.
For the initial vacuum state $\langle y \rangle=\langle p \rangle=0$ at all 
times, the (product of) Gaussians in amplitude space is the probability 
density $\rho(y_{01},~y_{02})$ with variances 
\begin{equation}
\sigma^2_{y_i}=\frac{1}{2}|f_k|^2=\Delta^2 y_i,~~~~~~~~~~i=1,2~, 
\end{equation}
while the Gaussians in momentum space are centered around 
$\frac{F(k)}{|f_k|^2}~y_i$ with variance $\sigma_{p_i}$ given by
\begin{equation}
\sigma^2_{p_i}=\frac{1}{8~|f_k|^2}\not=\Delta^2 p_i~~~~~~~~~~~~~~i=1,2~. 
\end{equation} 
We note that $\sigma^2_{p_i}$ is independent of $y_i$. From (\ref{wig}) it is 
natural to define $\Gamma$ as being of the order of the product of variances
\begin{equation}
\Gamma\sim \sigma_{y_1}~\sigma_{p_1}~\sigma_{y_2}~\sigma_{p_2} = \frac{1}{16},
\label{Gamma} 
\end{equation}
with some (unimportant) proportionality constant. Hence the 
typical volume occupied by the system in phase-space is invariant under 
squeezing, 
it is conserved and of the order of the minimal value of the quantity
$\Delta y_1~\Delta y_2~\Delta p_1~\Delta p_2$ allowed by the uncertainty 
principle. It is certainly negligible compared to $F^2(k)\gg 1$.
Note that this corresponds to $\Delta^2 y_1~\Delta^2{\overline p}_1=
\Delta^2 y_2~\Delta^2 {\overline p}_2=\frac{1}{4}$ where $y_{1,2},
{\overline p}_{1,2}\equiv 2p_{1,2}$ are canonically conjugate.
From the Wigner function given above we see that the Gaussian 
in $p$ space multiplying $|\Psi|^2$ must be interpreted as the conditional 
probability 
${\cal P}(p,~\eta~|~y,~\eta)$ introduced earlier. In the limit of infinite 
squeezing eq.~(\ref{Pcl}) is indeed recovered as is seen from (\ref{wig}).
\par
If we define now the entropy $S$ as we would do in classical statistical 
mechanics, we get 
\begin{equation}
S=-\int~\int~d\Gamma ~W~\ln (\alpha W) = \ln \Gamma +\beta,\label{S}
\end{equation}
with $d\Gamma\equiv 4~dy_{01}~dy_{02}~dp_{01}~dp_{02}$. It is seen 
that the entropy is the familiar expression in terms of $\Gamma$, it is 
constant under squeezing and can be put equal to zero by 
a proper choice of $\alpha$ as is desirable for any pure state. 
The entropy must remain zero under squeezing which   
just describes the unitary time evolution of our pure state. Therefore the 
idea to connect the entropy with the volume in phase-space using a formula 
like (\ref{S}) is fruitfull, however 
$\Gamma\not=\Delta p_1~\Delta y_1~\Delta p_2~\Delta y_2$.
\par
For $k\ll aH$, when the GW are in the quasi-static regime, the GW modes 
have the asymptotic expression 
\begin{equation} 
{\tilde f}_k = a~{\tilde \phi}_k=C_1~a + C_2~a\int_{\infty}^{\eta} 
a^{-2} d\eta'~,
~~~~~~~~~~~~~C_1~\Im C_2=-\frac{1}{2} \label{as}
\end{equation}
where $C_1$ can be made real and positive, both $C_1~{\rm and}~C_2$ are 
constants for a given wavelength. We have further
\begin{eqnarray}
P &=&\frac{1}{2f_{k1}}~e_p\\
&\stackrel{|r_k|\to \infty}{\longrightarrow}& g_{k_1}~e_p = 
-\frac{\Im C_2}{a}~e_p~.
\end{eqnarray}
The Wigner function then becomes
\begin{eqnarray}
W & \propto & |\Psi(y_0)|^2~\exp \left ( -\frac{|P_0|^2}{\Delta^2_P} \right)~.
\label{W}
\end{eqnarray}
A crucial point is that while
\begin{equation}
\Delta^2 y~\Delta^2 p \simeq f^2_{k1}~g^2_{k2}~\langle e^2_y \rangle^2
\gg 1~,~~~~~~~~~~~~~~k\ll aH,\label{yp} 
\end{equation}
we have on the other hand  
\begin{eqnarray}
\Delta^2 y~\Delta^2 P & \simeq & f^2_{k1}~g^2_{k1}~\langle e^2_y \rangle~
\langle e^2_p \rangle = a^4~\phi^2_{k_1}~\phi'^2_{k_2} \simeq 
\frac{1}{4}\label{yP1}\\
& \simeq & \Delta^2 {\tilde \phi}~\Delta^2 p_{\tilde \phi}\\
& \simeq & C^2_1~|\Im C_2|^2 
~~~~~~~~~~k\ll aH~.\label{yP2}
\end{eqnarray}
Note that $p_{\tilde \phi}$ is solely associated with the decaying mode. 
Equation (\ref{yp}) depends only on the growing mode and  
shows explicitly that $\Delta^2 y~\Delta^2 p$ cannot be a true measure 
for the volume occupied in phase-space and as said above the fact that this 
expression becomes very large has little to do with $\Gamma$. 
In contrast, equation (\ref{yP1}) makes use of both growing and 
decaying modes and is therefore a relevant quantity in this respect. 
The presence of the decaying mode 
may induce a quantum ``signature'' in the power spectrum $|f_k|^2$, as it 
avoids the presence of zeroes in the spectrum~\cite{david95}. 
However, this effect is very small since the spectrum is essentially 
determined by 
the growing mode. In contrast, the role 
played by the decaying mode here is crucial, once it is neglected the 
phase-space volume $\Gamma$ vanishes. If it is taken into account, $\Gamma$ 
is constant even though vanishingly small compared to $\Delta^2 y~\Delta^2 p$. 

\section{Summary and discussion}

We have shown that the typical phase-space volume $\Gamma$ of primordial GW 
generated during an inflationary stage remains finite under squeezing as seen 
in eq.(\ref{Gamma}) and we give an expression for it in the long wavelength 
regime, eqs (\ref{yP1}-\ref{yP2}).
It is enlightening to note in this connection that the canonical commutation 
relations just express Liouville's theorem on the conservation of volume in 
phase-space for the underlying classical system. Indeed classical GW will 
also satisfy 
\begin{equation}
g_k f^*_k + g^*_k f_k = 1~,
\end{equation}
a fact which just expresses Liouville's theorem on the  conservation 
of volume in phase space for Hamiltonian dynamics. Were the primordial GW 
classical and stochastic ab initio, they would also occupy 
a constant typical volume in phase space. For the primordial GW of quantum 
origin generated by inflation, the fact that 
$\Delta y~\Delta P$ remains finite and constant is just an expression 
of quantum coherence between the growing and decaying modes, so both of them 
are needed in order to determine the phase-space volume.
\par
It was further shown that the quasi-classical entropy can be defined as the 
logarithm of $\Gamma$ modulo a constant, eq.(\ref{S}) so that this entropy 
vanishes at all times for our pure initial vacuum state. 
Finally, we would like to stress the observational implications of the coarse 
graining procedure which is adopted for the calculation of the entropy. 
In a number of papers~\cite{pro,gasperini} the entropy of cosmological 
perturbations was calculated assuming that as a result of the interaction
of a quantum perturbation, which was initially in the pure vacuum state,
with the environment (or as a result of coarse graining), non-diagonal 
elements of the density matrix of the perturbation calculated in some 
particular basis quickly become zero. Formal neglection
of non-diagonal elements of the density matrix in both $N$-particles and
coherent states basis leads to the same value of the entropy
$S_k\simeq 2|r_k|$ for $|r_k|\gg 1$ for each mode ${\bf k}$. 
However, as a result of the procedure adopted, the correlation 
(\ref{pcl}) gets {\it completely} lost. In Fig.1, the neglection of 
non-diagonal elements corresponds to the replacement of the ellipse by a
large circle with radius equal to the major 
semi-axis of the ellipse. When (\ref{pcl}) holds, we have 
trajectories of measure zero in phase-space, so that the perturbations have 
a fixed temporal phase after the last Hubble radius crossing.
The latter is a crucial property. In the case of adiabatic perturbations, 
as is well known, it leads to the appearance of Sakharov oscillations in the 
power spectrum of matter and to multiple 
acoustic, or Doppler, peaks in the multipole spectrum $C_l$ of the 
microwave background temperature anisotropies at small angular scales. 
The latter are now the subject of 
intense investigation in connection with the future satellite missions, 
like COBRAS-SAMBA, which will be able to measure these peaks.

We believe that a more physical coarse graining should be used. It should
be based on a realistic model of the interaction of the perturbations
with the environment. A simple toy model is depicted in Fig.1
by the dashed circle which represents a stochastic GW background emitted by 
matter after the last Hubble radius crossing.
This background does not experience squeezing. Therefore, it has both 
stochastic amplitudes and stochastic phases. The radius of the dashed circle 
is much bigger than the radius of the thin solid circle. As a result, for the 
sum of the two GW backgrounds - squeezed primordial and non-squeezed 
secondary - the noise in {\em all} directions in phase-space turns out to be 
much bigger than for the vacuum state. Thus, as we pointed out in 
~\cite{david96}, such a state is not ``squeezed'' anymore from the 
observational point of view. On the other hand, the
radius of the dashed circle is much less than the major semi-axis of the
ellipse, so all predictions about {\it rms} values of perturbations and
their essentially fixed temporal phase remain unaltered. When applied to 
adiabatic perturbations, such a model does not destroy the multiple acoustic 
peaks of the $C_l$ multipoles. The entropy defined as the logarithm of the 
effective phase-space volume $\Gamma$ for the sum of these two 
backgrounds, though non-zero anymore, is still much smaller than $2|r_k|$ as 
is clear from Fig.1. Work on this issue is currently under progress. 

\vspace{1cm}
\noindent
{\bf Acknowledgements}
\par\noindent
A significant part of this project was accomplished during the visit of
one of the authors (A.S.) to France under
the agreement between the Landau Institute for Theoretical Physics and
Ecole Normale Sup\'erieure, Paris. A.S. thanks ENS and EP93 CNRS (Tours) for
financial support and Profs E. Brezin, C. Barrabes for their hospitality
in ENS, Universit\'e de Tours respectively. A.S. also acknowledges financial
support by the Russian Foundation for Basic Research, grant 96-02-17591,
and by Russian research project ``Cosmomicrophysics''.

\end{document}